\documentclass[11pt]{article}
\usepackage{mathrsfs}
\usepackage{graphicx}
\begin{document}
%\begin{CJK*}{GBK}{song}

\title{Topological superfluid of spinless Fermi gases in p-band honeycomb optical lattices with on-site rotation}

\author{Beibing Huang \\
Department of Experiment Teaching, Yancheng Institute of Technology,
Yancheng, 224051, China
\\Xiaosen Yang and ShaoLong Wan\\
Institute for Theoretical Physics and Department of Modern Physics \\
University of Science and Technology of China, Hefei, 230026, China}

\maketitle
\begin{center}
\begin{minipage}{120mm}
\vskip 0.8in
\begin{center}{\bf Abstract} \end{center}

{In this paper, we put forward to another route realizing
topological superfluid (TS). In contrast to conventional method,
spin-orbit coupling and external magnetic field are not requisite.
Introducing an experimentally feasible technique called on-site
rotation (OSR) into p-band honeycomb optical lattices for spinless
Fermi gases and considering CDW and pairing on the same footing, we
investigate the effects of OSR on superfluidity. The results suggest
that when OSR is beyond a critical value, where CDW vanishes, the
system transits from a normal superfluid (NS) with zero TKNN number
to TS labeled by a non-zero TKNN number. In addition, phase
transitions between different TS are also possible.}

\end{minipage}
\end{center}

\vskip 1cm

\textbf{PACS} number(s): 67.85.Lm, 03.65.Vf, 74.20.-z

\section{Introduction}

Topological superfluid  (or superconductor) (TS) has a full pairing
gap in the bulk and is labeled by a non-zero integer topological
invariant \cite{hasan, qi}. From the famous bulk-boundary
correspondence such a topological integer ensures the existence of
gapless excitations on the boundary of the system, in other words
Majorana fermions (MF) \cite{majorana} in vortex core of pairing
order parameter. Roughly, MFs are neither fermions nor bosons but
non-Abelian anyons \cite{ia} and play an important role for the
realization of fault-tolerant topological quantum computation (TQC)
\cite{sarma}. The application prospect of MFs makes TS become one of
the hottest frontiers.

In the condensed matter physics some practical two-dimensional
systems have been theoretically proposed to realize TS \cite{read,
tsut, sato1, lee, fu, linder, jau, anj1}. In terms of these systems
the entrance into TS requires subtle adjustment of Hamiltonian and
it is very difficult in condensed matter physics, although MFs have
been detected in InSb nanowires contacted with one normal (Au) and
one superconducting electrode (NbTiN) \cite{rea}. In the light of
the disadvantage for condensed matter, TS has been also suggested in
cold Fermi gases owing to their many controllable advantages and
operabilities. Following the successful observation p-wave Feshbach
resonance (FR), Gurarie et al. \cite{gurarie} show that degenerate
Fermi gases near a p-wave FR naturally give a concrete realization
of TS. Zhang et al. \cite{zhang} propose to create TS directly from
an s-wave interaction making use of an artificially generated
spin-orbit coupling (SOC). In fact, SOC have been realized in a
neutral atomic Bose-Einstein condensate (BEC) by dressing two atomic
spin states with a pair of lasers and the same technique is also
feasible for cold Fermi gases \cite{e1, e2}. Realizing that in a
dual transformation SOC is formally equivalent to a p-wave
superfluid gap, Sato et al. \cite{sato} suggest to artificially
generate the vortices of SOC by using lasers carrying orbital
angular momentum. In terms of the latter two ways, SOC and a large
magnetic field are crucial in order to enter into TS.

In this paper we suggest to create TS from spinless Fermi gases in
p-band honeycomb optical lattices with so-called on-site rotation
(OSR), that rotates every lattice site around its own center but
keeps the whole lattice intact and has been realized for triangular
optical lattices \cite{gem}. As a matter of fact p-band Fermi gases
in honeycomb optical lattices in absence of OSR have shown many
interesting characteristics, such as ferromagnetism \cite{ferr} and
Wigner crystallization \cite{wigner, wigner1} associating with flat
bands, f-wave superfluidity with conventional pairing interaction
\cite{fwave}. The motivation for this paper comes from the Wu's work
on quantum anomalous Hall effect in the same system \cite{wu}. Under
single particle picture, Wu found that an arbitrary non-zero OSR not
only breaks time-reversal symmetry and changes the topological
properties of the system, but also drives a topological phase
transition when OSR is beyond a critical value. Here we add on-site
attraction interaction between p-band Fermi atoms into Hamiltonian
and ask a question whether OSR can drive a phase transition into TS.
The results are positive and OSR brings phase transitions not only
from normal superfluid (NS) to TS, but also among different TS. From
another perspective our work also can be considered as an extension
to \cite{fwave}, where f-wave superfluidity without OSR is
discussed. Thus we also investigate the effects of OSR on f-wave
superfluidity.

Experimentally the route to realize TS suggested here is also
feasible. On the one hand by placing two electro-optic modulators at
two of three laser beams which coherently superpose to form a
honeycomb lattice, OSR is available as illustrated in \cite{machi}.
On the other hand due to Pauli exclusion principle the occupation of
p-band is very convenient as long as the lowest s-band is fulfilled.
In addition, on-site attraction interaction can be enhanced by using
atoms with large magnetic moments, such as ${}^{167}$Er with $m =
7\mu_B$ on which laser cooling has been performed \cite{35}. In
contrast to \cite{zhang, sato}, where a pair of extra lasers and a
large magnetic field are needed to produce an effective SOC and
split two SOC bands respectively, our system is much simpler.

The organization of this paper is as follows. In section 2, we give
the model and at the mean-field level investigate the ground state
of the system by numerically minimizing the thermodynamic potential.
In section 3 by calculating TKNN number $I_{TKNN}$ \cite{TKNN} of
occupied bands addressing the topological properties of the model,
the topological phase diagram is obtained. In addition we also
investigate the properties of edge states to prove our results. A
brief conclusion is given in section 4.

\section{Model and Mean-Field Ground State}

The honeycomb optical lattice was realized experimentally by using
three laser beams with co-planar propagating wavevectors quite some
time age \cite{salomon}. It is well known that a honeycomb lattice
is not a Bravais lattice and there are two inequivalent sites in a
unit cell, denoted by A and B respectively. Fulfilling the lowest
s-band and defining three unit vectors
$\vec{e}_1=\frac{\sqrt{3}}{2}\vec{e}_x+\frac{1}{2}\vec{e}_y$,
$\vec{e}_2=-\frac{\sqrt{3}}{2}\vec{e}_x+\frac{1}{2}\vec{e}_y$ and
$\vec{e}_3=-\vec{e}_y$, the Hamiltonian of p-band honeycomb optical
lattices with OSR is
\begin{eqnarray}
H=t_{\|}\sum_{\vec{r}\in A, i}\left[p_{\vec{r},
i}^{\dag}p_{\vec{r}+\vec{e}_i, i}+H.C.\right]- U\sum_{\vec{r}\in
A\oplus
B}p_{\vec{r}x}^{\dag}p_{\vec{r}y}^{\dag}p_{\vec{r}y}p_{\vec{r}x}-\Omega
\sum_{\vec{r}\in A\oplus B}\hat{l}_{\vec{r}, z}-\mu\sum_{\vec{r}\in
A\oplus B}\hat{n}_{\vec{r}}, \label{1}
\end{eqnarray}
where $p_{\vec{r}, i}=(p_{\vec{r}, x} \vec{e}_x+p_{\vec{r}, y}
\vec{e}_y)\cdot\vec{e}_i$ and $p_{\vec{r}, x}$ ($p_{\vec{r}, y}$) is
the annihilation operator for $p_x$ ($p_y$) band at the lattice site
$\vec{r}$. $\hat{n}_{\vec{r}}=p_{\vec{r}, x}^{\dag} p_{\vec{r}, x}+
p_{\vec{r}, y}^{\dag} p_{\vec{r}, y}$ and $\hat{l}_{\vec{r},
z}=-i(p_{\vec{r}, x}^{\dag} p_{\vec{r}, y}- p_{\vec{r}, y}^{\dag}
p_{\vec{r}, x})$ represent particle number and orbital angular
moment operators. $t_{\|}$ is the nearest-neighbor hopping matrix
element of atoms in $\sigma$ bonds and positive due to the odd
parity of the p-orbital. $U$ ($>0$), $\Omega$ ($>0$) and $\mu$ are
the on-site interaction strength, on-site rotation angular velocity
and chemical potential, respectively. Note that we have neglected
the nearest-neighbor atom hopping of $\pi$ bonds and supposed the
nearest neighbor distance in the lattice to be unit.

When $U=0$, introducing operator $\phi(k)=[p_{Ax}(k), p_{Ay}(k),
p_{Bx}(k), p_{By}(k)]^T$ and making a unitary transformation
$\phi_n(k)=U_{nm}(k)\Psi_m(k)$, Hamiltonian can be diagonalized
exactly. Meanwhile four energy bands can be obtained. Wu found two
of four bands always are topological for any nonzero OSR and the
others can be topological only if OSR is beyond a critical value
\cite{wu}. On the basis of this findings, Wu proposed an orbital
analogue of the quantum anomalous Hall effect, arising from orbital
angular momentum polarization due to OSR. With $\Omega=0$, Lee et
al. discussed f-wave superfluidity and charge density wave (CDW) in
this system at the mean-field level \cite{fwave}. Their results show
that away from the half filling the system is f-wave superfluidity,
while around the half filling superfluidity and CDW coexist and the
system is a supersolid. Although superfluidity exists all the time,
it is not topological as stated below.

Following the same spirit in \cite{fwave} we decouple interaction
term into CDW channel
\begin{eqnarray}
H_{int}^{CDW} =\sum_{\tau=x,y}\left[\sum_{\vec{r}\in
A}(-\frac{n}{2}U-\frac{\Delta_{CDW}}{2})p_{\vec{r}, \tau}^{\dag}
p_{\vec{r}, \tau} + \sum_{\vec{r}\in
B}(-\frac{n}{2}U+\frac{\Delta_{CDW}}{2})p_{\vec{r}, \tau}^{\dag}
p_{\vec{r}, \tau}\right] \label{2}
\end{eqnarray}
and pairing channel
\begin{eqnarray}
H_{int}^{pairing}&=&-\sum_k\left[\Delta_A p_{Ax}^{\dag}(k)
p_{Ay}^{\dag}(-k) + \Delta_B p_{Bx}^{\dag}(k) p_{By}^{\dag}(-k) +
H.C. \right] \nonumber\\
&=& -\sum_{k'}\left[\Delta_{nm}(k')\Psi_{n}^{\dag}(k')
\Psi_{m}^{\dag}(-k') + H.C.\right] \label{3}
\end{eqnarray}
where $n=<\hat{n}_{\vec{r}_A}+\hat{n}_{\vec{r}_B}>/2$ is filling
factor of every site,
$\Delta_{CDW}=U<\hat{n}_{\vec{r}_A}-\hat{n}_{\vec{r}_B}>/2$,
$\Delta_{A}=U\sum_k <p_{Ay}(-k) p_{Ax}(k)>$, $\Delta_{B}=U\sum_k
<p_{By}(-k) p_{Bx}(k)>$ are order parameters for CDW and
superfluidity. In (\ref{3}) we also express the pairing channel
using quasiparticle $\Psi(k)$. In this representation
\begin{eqnarray}
\Delta_{nm}(k)&=&\Delta_A\left[ U_{1n}^{\ast}(k)U_{2m}^{\ast}(-k) -
U_{2n}^{\ast}(k)U_{1m}^{\ast}(-k)\right]\nonumber\\ &+&
\Delta_B\left[ U_{3n}^{\ast}(k)U_{4m}^{\ast}(-k) -
U_{4n}^{\ast}(k)U_{3m}^{\ast}(-k)\right]. \label{4}
\end{eqnarray}

After the mean-field approximation, the Hamiltonian (\ref{1})
becomes a BdG Hamiltonian $H=[\phi^{\dag}(k), \phi(-k)] H_k
[\phi(k), \phi^{\dag}(-k)]^T$ and the properties of system are
completely decided by the $8\times 8$ matrix $H_k$. Diagonalizing
$H_k$, we attain spectrum $\epsilon_i(k)$ and correspondingly
eigenvectors $\varphi_i(k)$ ($i=1,2,\cdot\cdot\cdot, 8$). Due to
particle-hole symmetry inherent in this BdG Hamiltonian, the
spectrum are symmetric about zero energy and we assume
$\epsilon_1(k)=-\epsilon_8(k)>0$, $\epsilon_2(k)=-\epsilon_7(k)>0$,
$\epsilon_3(k)=-\epsilon_6(k)>0$, $\epsilon_4(k)=-\epsilon_5(k)>0$.
Then the thermodynamical potential at zero temperature is
\begin{eqnarray}
F=\frac{1}{2}\sum_k
\left[-4\mu-\epsilon_1(k)-\epsilon_2(k)-\epsilon_3(k)-\epsilon_4(k)\right]+\frac{N}{U}
|\Delta_A|^2+ \frac{N}{U} |\Delta_B|^2+\frac{N}{2U} \Delta_{CDW}^2,
\label{5}
\end{eqnarray}
where $N$ is the number of the unit cell. Below we numerically
minimize thermodynamic potential $F$ about $\Delta_A$, $\Delta_B$
and $\Delta_{CDW}$ for fixed interaction strength $U$. Without loss
of generality we choose $\Delta_A$ to be real,
$\Delta_B=|\Delta_B|e^{i\theta}$ and $U/t_{\|}=3.0$.

Fig.1 shows the solutions of the Hamiltonian (\ref{1}) at the
mean-field level for changing chemical potential $\mu$ and OSR
$\Omega$. Due to the particle-hole symmetry we only concentrate on
negative chemical potential. Fig.1(a) describes the variation of
$\Delta_{CDW}$. For $\Omega=0$ CDW is robust, but when $\Omega$ is
beyond a critical value $\Omega_c$ it vanishes suddenly. This is due
to the fact that the appearance of OSR changes the band structures
of single particle and breaks the nesting condition for CDW.
Numerically we find $\Omega_c/t_{\|}\approx 0.4\sim 0.6$ and is
monotonically increasing as the function of chemical potential.
Fig.1(b) shows the effect of OSR on particle density and further
exemplifies that the variations of band structures driven by OSR
cause nonmonotonic behavior of particle density. In contrast,
superfluid order parameters $\Delta_{A}$, $\Delta_{B}$ are more
interesting and shown in (c) and (d). On the one hand for
$\Omega>\Omega_c$, $\Delta_{A}=|\Delta_{B}|$ and with the increase
of OSR superfluid order smoothly decreases until disappearance. This
suppression mechanism of superfluidity consists in time-reversal
symmetry broken caused by OSR. While on the other hand for
$\Omega<\Omega_c$ $\Delta_{A}$ is still decreasing but $\Delta_{B}$
is increasing with $\Omega$. In fact the increase of $\Delta_{B}$
originates from the redistribution of particle density between sites
A and B, in other words the decrease of $\Delta_{CDW}$ as seen in
(a). Thus at the mean-field level our calculation suggests that (1)
OSR weakens stabilities of CDW and superfluidity and (2) for
$\Omega<\Omega_c$, superfluidity and CDW coexist and the system is a
supersolid.

The optimization of $\theta$ leads to $\theta=\pi$ for all
parameters we choose. Below we discuss the effects of OSR on pairing
symmetry for $\Omega>\Omega_c$. From \cite{fwave} without OSR and
away from the half filling the intraband pairings in (\ref{4}) have
f-wave symmetry with three nodal lines of $k_x = 0$, $k_y = \pm
k_x/\sqrt{3}$ and $\pi/3$ rotation symmetry [Fig.2(d)]. On
introducing OSR, in terms of the pairing magnitude, nodal lines
degenerate into some disconnected regions where intraband gap
disappears, and $\pi/3$ rotation symmetry retains. However real and
imaginary parts of pairing break $\pi/3$ into $\pi$ rotation
symmetry. In Fig.2(a) (b) and (c) as an example we show the
magnitude, real and imaginary parts of $\Delta_{11}$.

\section{Topological Phase Diagram and Majorana Fermion Modes}

In this section we discuss topological properties of Hamiltonian
(\ref{1}). In terms of our system, it explicitly breaks the
time-reversal symmetry due to OSR. Thus TKNN number $I_{TKNN}$ plays
a central role in deciding topological nature of the system
\cite{TKNN}. TKNN number is defined, by eigenvectors $\varphi_i(k)$
($i=5,6,7,8$) corresponding to negative energy spectrm of the matrix
$H_k$, into $ I_{TKNN}=\frac{1}{2\pi i}\int d^2k\, Tr\, dA$, where
$A$ is a matrix one-form $A_{ij}=A_{ij}^{\nu}(k)dk_{\nu}$ with
$A_{ij}^{\nu}(k)=\varphi_i^{\dag}(k) \nabla_{k_{\nu}}\varphi_j(k)$.
By numerically calculating TKNN number \cite{fukui}, we show the
topological phase diagram of the system in Fig.3. For parameter
region we choose, there are four different subregions labeled by
$I_{TKNN}=1,0,-1,2$ respectively. Moreover by comparison with
Fig.1(a) it is easily found that the boundary between $I_{TKNN}=0$
and other TKNN numbers in the direction of $\Omega$ coincides with
that of CDW disappearance. This finding is very important and
ensures that topological order of our system is not topological CDW
\cite{satom}. According to the criteria for TS \cite{satom}
$I_{TKNN}=2$ corresponds to Abelian TS while $I_{TKNN}=1,-1$ are
non-Abelian TS. Thus Fig.3 tells us that OSR drives topological
phase transition not only from NS to TS, but also between different
TS. Here we mention a fact the energy gap of the bulk spectrum
closes when topological phase transitions between topologically
distinct phases occur.

From the bulk-edge correspondence, a non-trivial bulk topological
number implies the existence of gapless edge states localized on
open edges of the system. Cold Fermi gases with sharp edges may be
realized along the lines proposed in \cite{goldman}. In order to
understand the relation between $I_{TKNN}$ and the number of edge
states, we study the Hamiltonian (\ref{1}) with the open boundary
condition along the zigzag edge of the honeycomb lattice. The
resulting excitation spectrum are depicted in Fig.4 for
representative parameter choices. Very explicitly the number of
gapless states for every edge is one-to-one correspondence with the
TKNN number. For $I_{TKNN}=\pm 1$ ($I_{TKNN}= 2$) there are one
(two) pair(s) of gapless states, while for $I_{TKNN}=0$, gapless
state does not exist. Due to particle-hole symmetry, in terms of
gapless states, they are Majorana fermion modes. It should also be
remembered that the core of a vortex is topologically equivalent to
an edge which has been closed on itself. The edge modes we describe
are therefore equivalent to the Majorana fermions known to exist in
the core of vortices of p-wave superfluids \cite{lewenstein}.

\section{Conclusions}

In conclusion at the mean-field level we have investigated the
effects of OSR on CDW and superfluidity for p-band spinless Fermi
gases in honeycomb optical lattices. We found that OSR weakens the
stabilities of CDW and superfluidity simultaneously, although
superfluidity can survives a larger OSR. This conclusion leads to
another important result that once CDW drops out the system enters
into topological superfluidity. By numerically calculating the TKNN
number we obtained topological phase diagram of the system. In
addition edge states, i.e. bulk-boundary correspondence are also
investigated.

\section*{Acknowledgement}

The work was supported by National Natural Science Foundation of
China under Grant No. 10675108 and Foundation of Yancheng Institute
of Technology under Grant No. XKR2010007.

\begin{figure}[htbp]
\includegraphics[width=7.5cm, height=6.0cm]{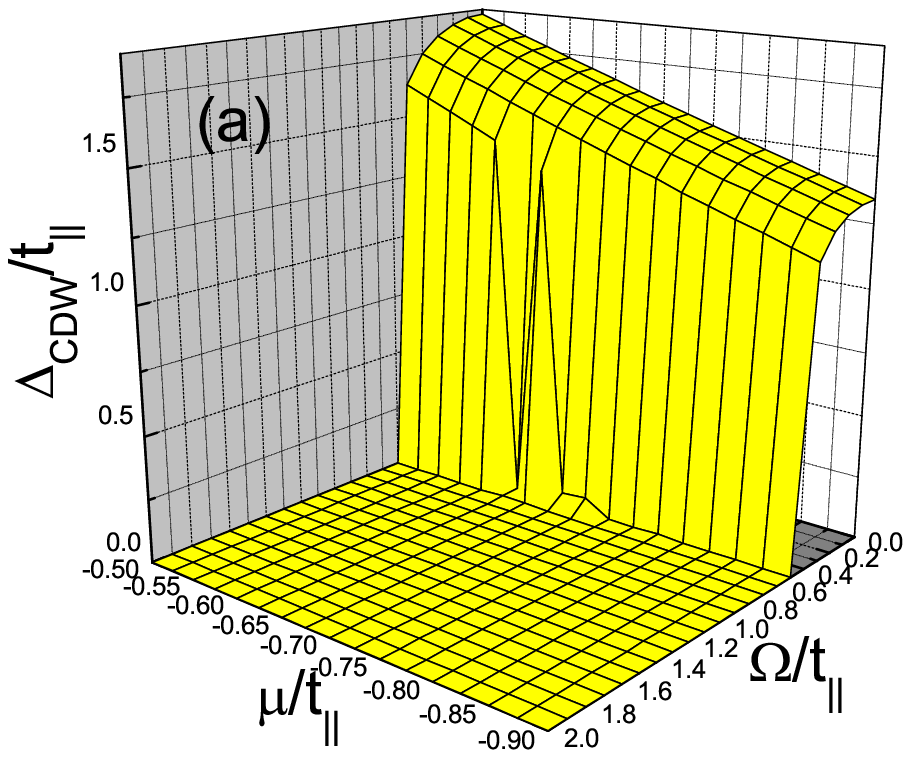}
\includegraphics[width=7.5cm, height=6.0cm]{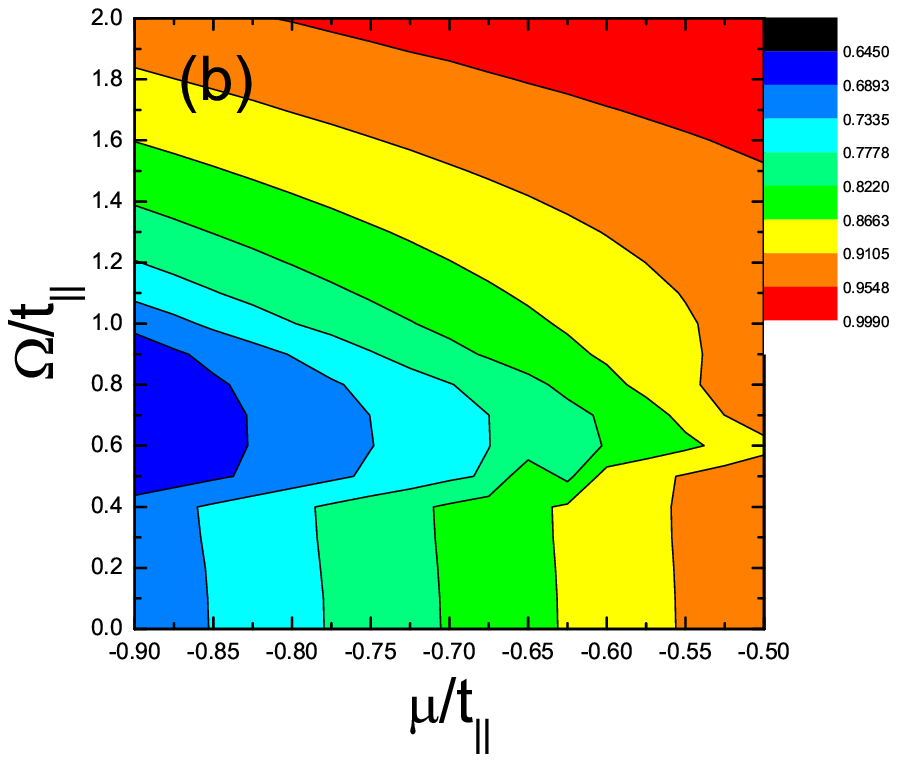}
\includegraphics[width=7.5cm, height=6.0cm]{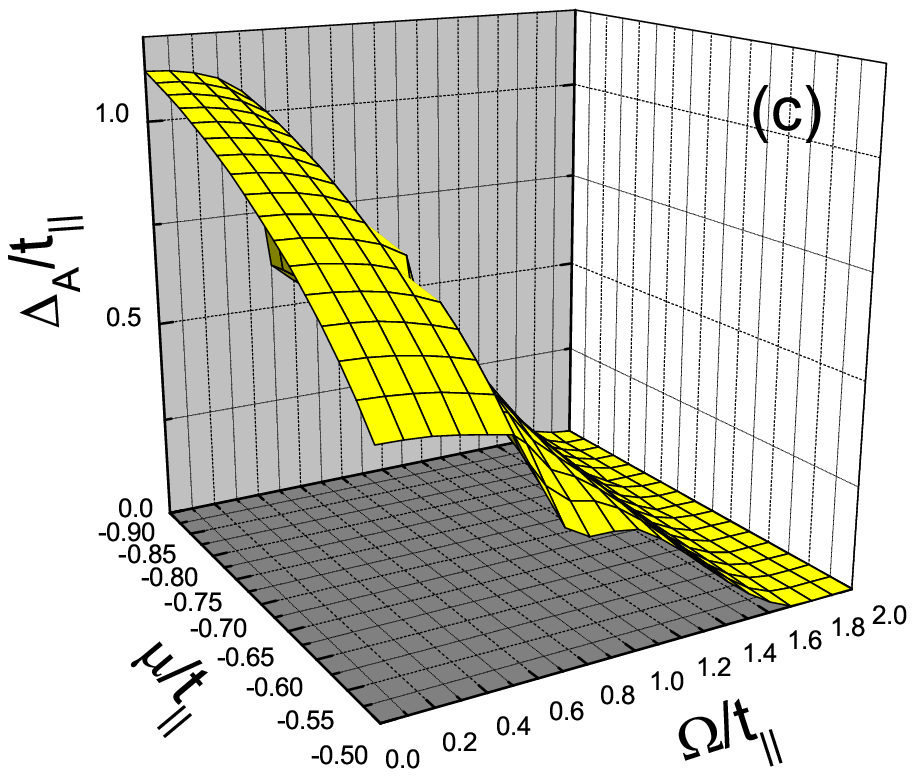}
\includegraphics[width=7.5cm, height=6.0cm]{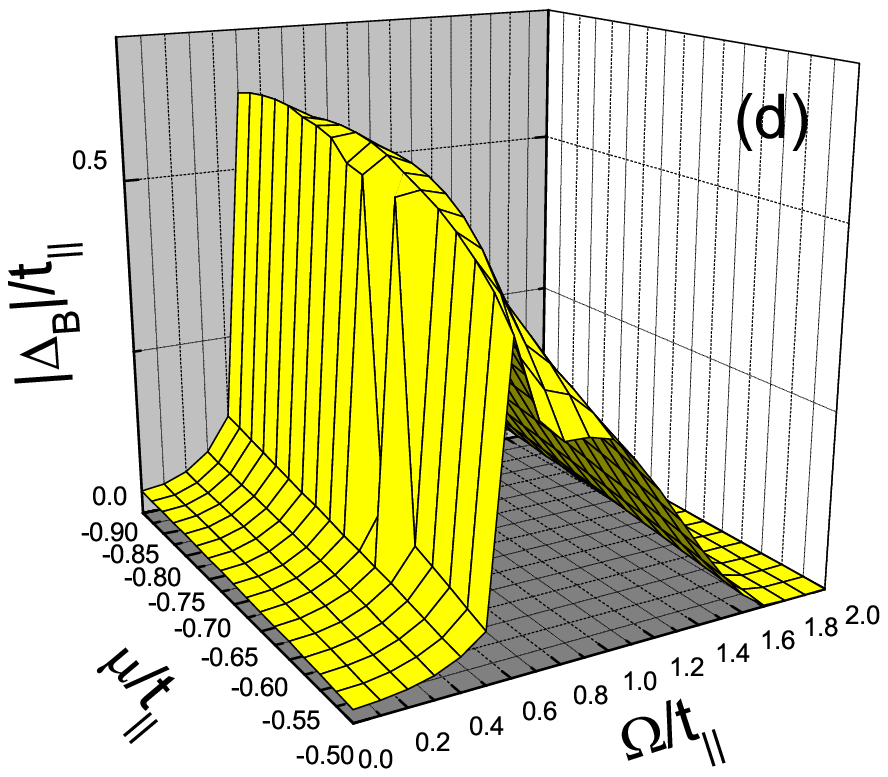}
\caption{The mean-field solution of the Hamiltonian (\ref{1}).
Parameter $U/t_{\|}=3.0$.} \label{fig.1}
\end{figure}

\begin{figure}[htbp]
\includegraphics[width=7.5cm, height=6.0cm]{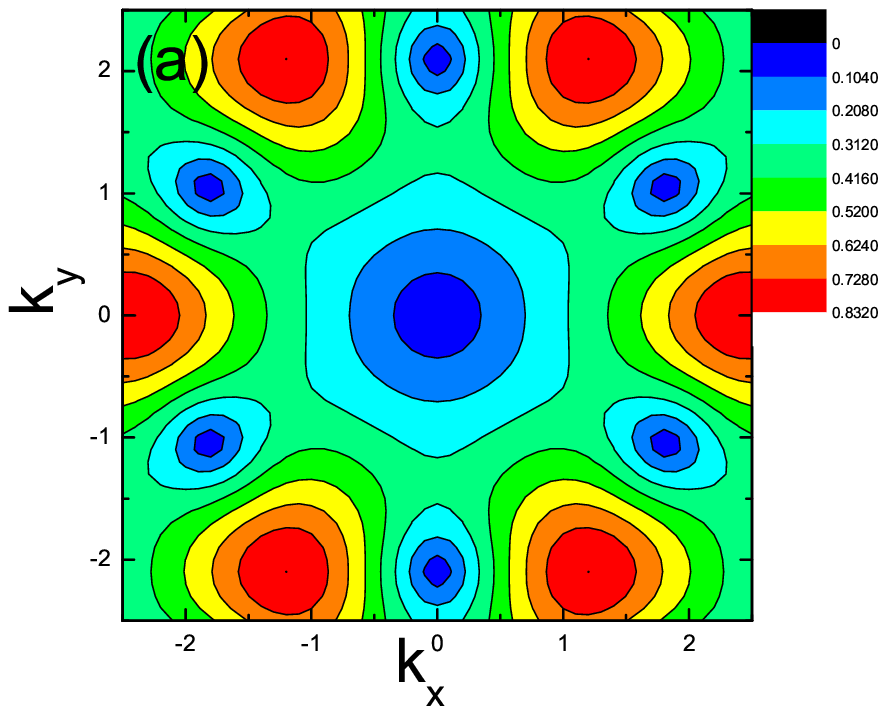}
\includegraphics[width=7.5cm, height=6.0cm]{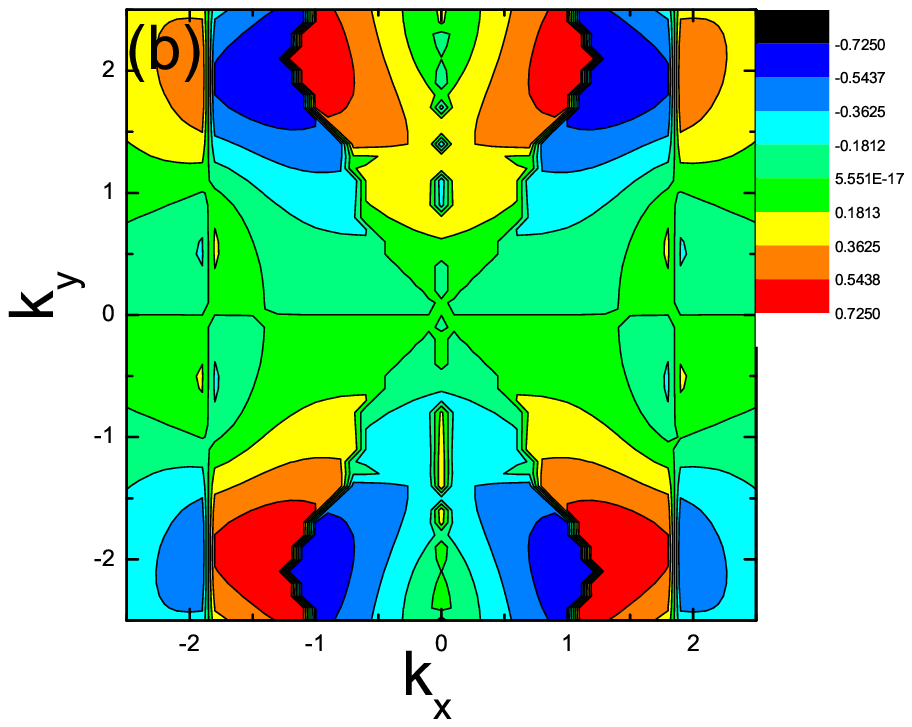}
\includegraphics[width=7.5cm, height=6.0cm]{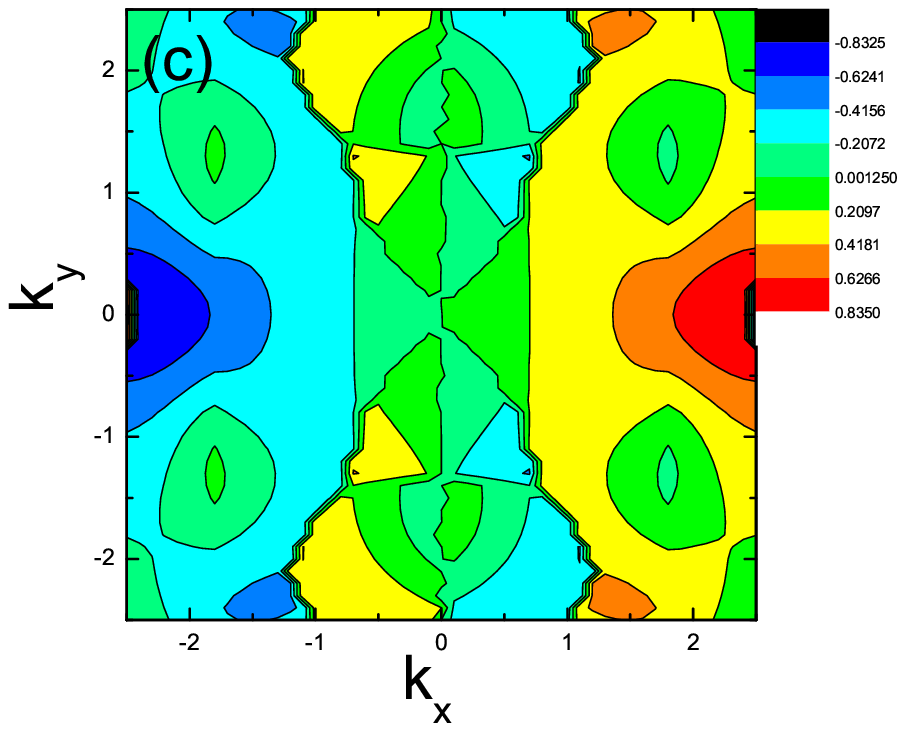}
\includegraphics[width=7.5cm, height=6.0cm]{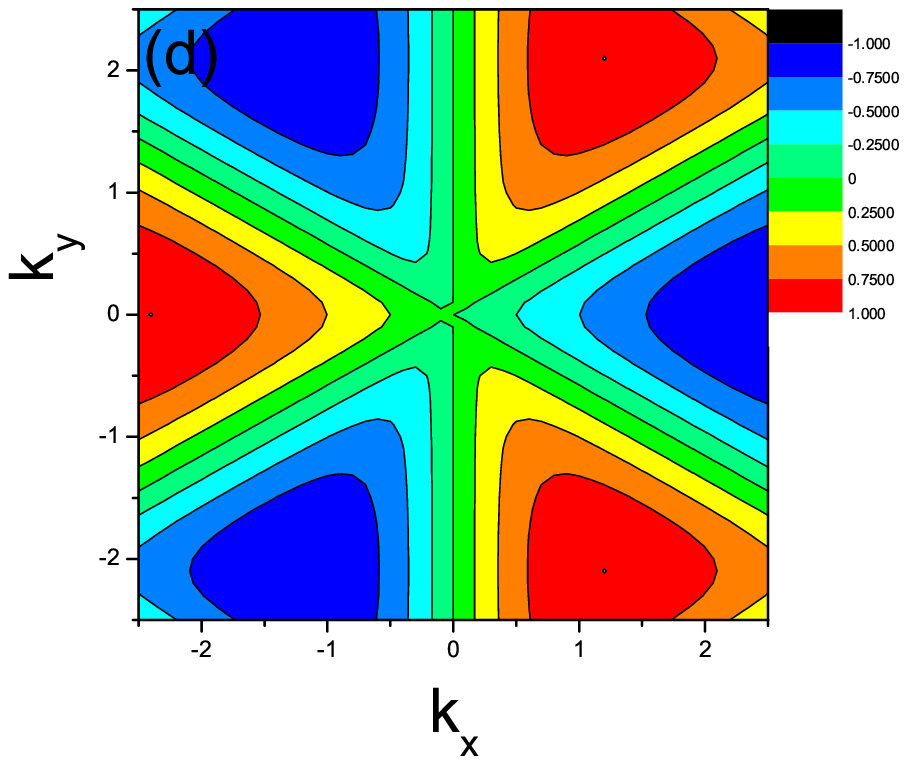}
\caption{The symmetry of intraband pairing $\Delta_{11}$. In (a) the
magnitude, (b) real part and (c) imaginary part of $\Delta_{11}$ for
$\Omega/t_{\|}=1.0$ are shown. For comparison (d) plots
$\Delta_{11}$ for $\Omega/t_{\|}=0$.} \label{fig.2}
\end{figure}

\begin{figure}[htbp]
\includegraphics[width=7.5cm, height=6.0cm]{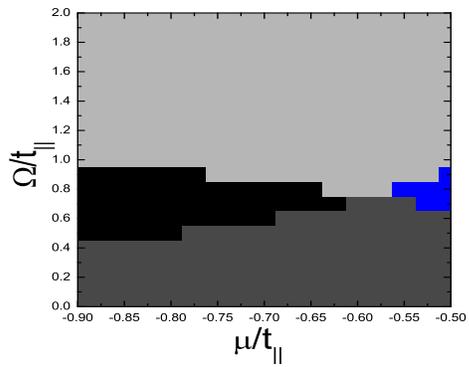}
\caption{Topological phase diagram of the Hamiltonian (\ref{1}) at
the mean-field level. The light grey, dark grey, black and blue
colors correspond to $I_{TKNN}=1,0,-1,2$ respectively. Parameter
$U/t_{\|}=3.0$.} \label{fig.3}
\end{figure}

\begin{figure}[htbp]
\includegraphics[width=7.5cm, height=6.0cm]{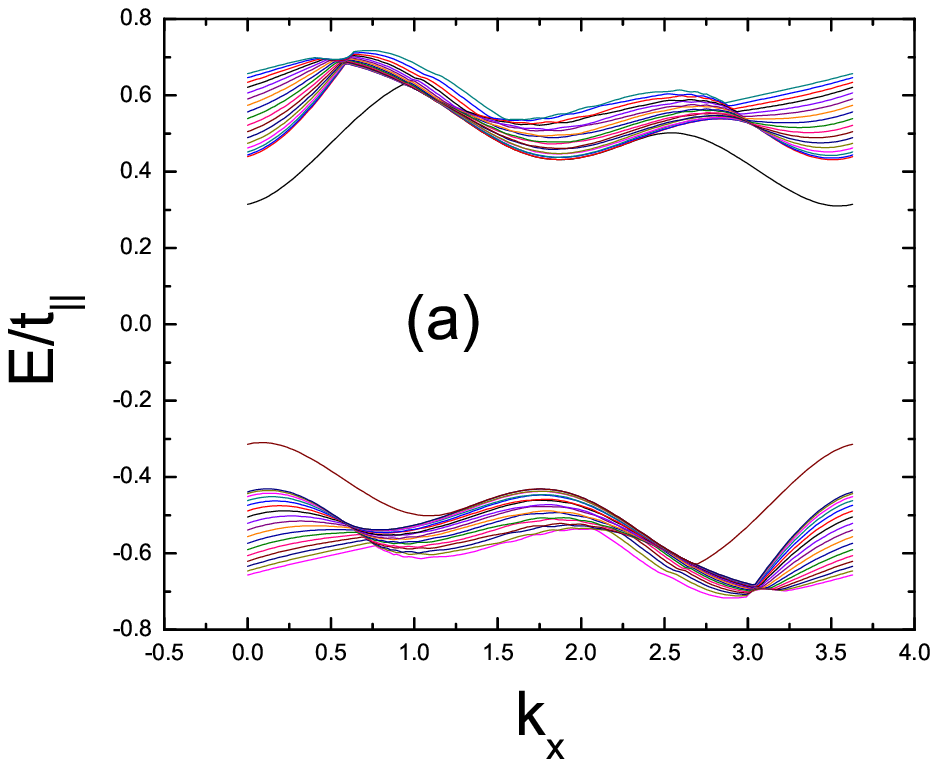}
\includegraphics[width=7.5cm, height=6.0cm]{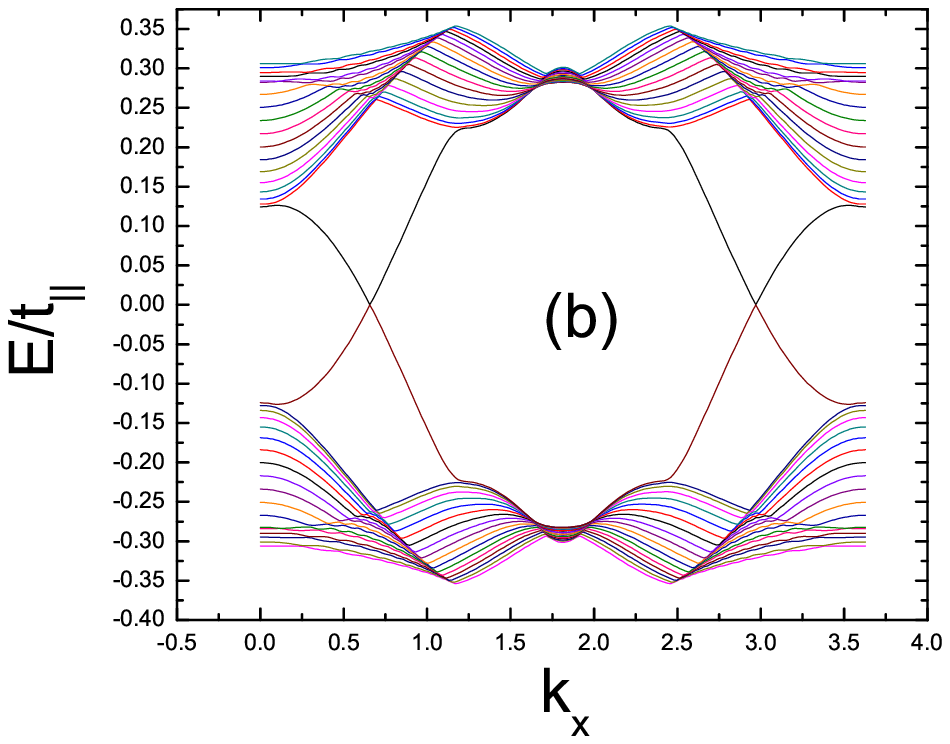}
\includegraphics[width=7.5cm, height=6.0cm]{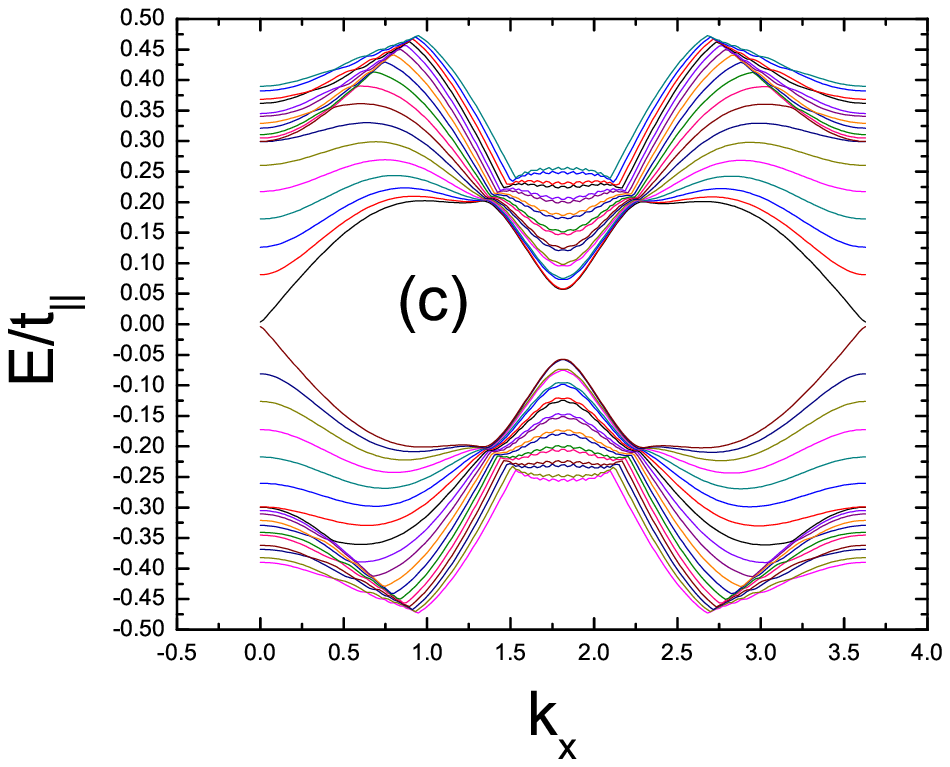}
\includegraphics[width=7.5cm, height=6.0cm]{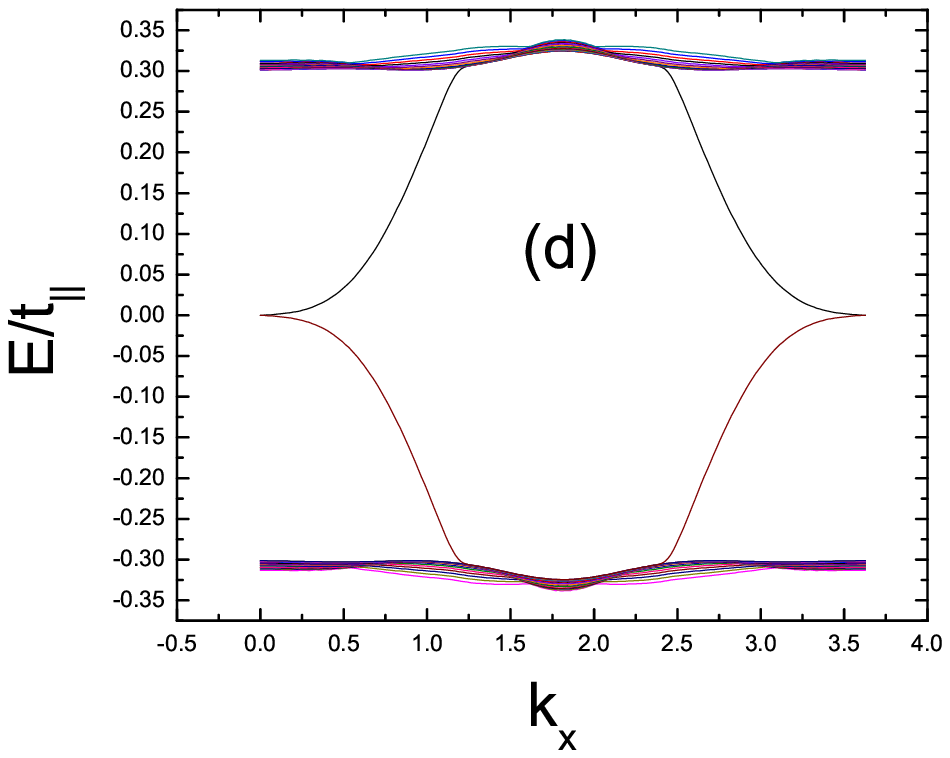}
\caption{The gapless edge states with the open boundary condition
along the zigzag edge of the honeycomb lattice. In (a) $I_{TKNN}=0$,
$\mu/t_{\|}=-0.75$, $\Omega/t_{\|}=0.3$, $\Delta_A/t_{\|}=0.969$,
$\Delta_B/t_{\|}=0.063$, $\Delta_{CDW}/t_{\|}=1.462$, (b)
$I_{TKNN}=2$, $\mu/t_{\|}=-0.5$, $\Omega/t_{\|}=0.8$,
$\Delta_A/t_{\|}=\Delta_B/t_{\|}=0.288$, $\Delta_{CDW}/t_{\|}=0$,
(c) $I_{TKNN}=-1$, $\mu/t_{\|}=-0.85$, $\Omega/t_{\|}=0.8$,
$\Delta_A/t_{\|}=\Delta_B/t_{\|}=0.523$, $\Delta_{CDW}/t_{\|}=0$,
(d) $I_{TKNN}=1$, $\mu/t_{\|}=-0.65$, $\Omega/t_{\|}=1.1$,
$\Delta_A/t_{\|}=\Delta_B/t_{\|}=0.307$, $\Delta_{CDW}/t_{\|}=0$.}
\label{fig.4}
\end{figure}

%\end{CJK*}

\begin{thebibliography}{99}
\bibitem{hasan} M. Z. Hasan and C. L. Kane, Rev. Mod. Phys. 82,
3045 (2010).
\bibitem{qi} X.-L. Qi and S.-C. Zhang, Rev. Mod. Phys. 83,
1057 (2011).
\bibitem{majorana} E. Majorana, Nuovo Cimento 5, 171 (1937).
\bibitem{ia} D. A. Ivanov, Phys. Rev. Lett. 86, 268 (2001).
\bibitem{sarma} C. Nayak, S. H. Simon, A. Stern, M. Freedman and S. Das Sarma,
Rev. Mod. Phys. 80, 1083 (2008).
\bibitem{read} N. Read and D. Green, Phys. Rev. B 61, 10267 (2000).
\bibitem{tsut} Y. Tsutsumi, T. Kawakami, T. Mizushima, M. Ichioka and
K. Machida, Phys. Rev. Lett. 101, 135302 (2008).
\bibitem{sato1} M. Sato and S. Fujimoto, Phys. Rev. B 79, 094504 (2009).
\bibitem{lee} P. A. Lee, arXiv:0907.2681.
\bibitem{fu} L. Fu and C. L. Kane, Phys. Rev. Lett. 100, 096407 (2008).
\bibitem{linder} J. Linder, Y. Tanaka, T. Yokoyama, A. Sudbo and N.
Nagaosa, Phys. Rev. Lett. 104, 067001 (2010).
\bibitem{jau} J. D. Sau, R. M. Lutchyn, S. Tewari, and S. Das Sarma, Phys. Rev.
Lett. 104, 040502 (2010).
\bibitem{anj1} J. Alicea, Phys. Rev. B 81, 125318 (2010).
\bibitem{rea} V. Mourik, K. Zuo, S. M. Frolov, S. R. Plissard, E. P. A. M. Bakkers
and L. P. Kouwenhoven, Science 336, 1003 (2012).
\bibitem{gurarie} V. Gurarie, L. Radzihovsky and A. V. Andreev, Phys. Rev. Lett.
94, 230403 (2005).
\bibitem{zhang} C. Zhang, S. Tewari, R.M. Lutchyn and S. Das Sarma, Phys. Rev.
Lett. 101, 160401 (2008).
\bibitem{e1} P. Wang, Z. Yu, Z. Fu, J. Miao, L. Huang, S. Chai, H. Zhai and J.
Zhang, arXiv:1204.1887.
\bibitem{e2} L. W. Cheuk, A. T. Sommer, Z. Hadzibabic,
T. Yefsah, W. S. Bakr and M. W. Zwierlein, arXiv:1205.3483.
\bibitem{sato} M. Sato, Y. Takahashi and S. Fujimoto, Phys. Rev. Lett. 103,
020401 (2009).
\bibitem{gem} N. Gemelke, E. Sarajlic, S. Chu, arXiv:1007.2677.
\bibitem{ferr} S. Zhang, H. Hung and C. Wu, Phys. Rev. A 82, 053618
(2010).
\bibitem{wigner} C. Wu, D. Bergman, L. Balents and S. Das Sarma, Phys.
Rev. Lett. 99, 070401 (2007).
\bibitem{wigner1} C. Wu and S. Das Sarma, Phys.
Rev. B 77, 235107 (2008).
\bibitem{fwave} W. Lee, C. Wu and S. Das Sarma, Phys. Rev. A 82, 053611
(2010).
\bibitem{wu} C. Wu, Phys. Rev. Lett. 101, 186807 (2008).
\bibitem{machi} M. Zhang, H. Hung, C. Zhang and C. Wu, Phys. Rev. A 83, 023615 (2011).
\bibitem{35} J. J. McClelland and J. L. Hanssen, Phys. Rev. Lett.
96, 143005 (2006).
\bibitem{TKNN} D. J. Thouless, M. Kohmoto, M. P. Nightingale and M. Nijs, Phys. Rev. Lett. 49,
405 (1982).
\bibitem{salomon} G. Grynberg, B. Lounis, P. Verkerk, J.-Y. Courtois and C.
Salomon, Phys. Rev. Lett. 70, 2249 (1993).
\bibitem{fukui} T. Fukui, Y. Hatsugai and H. Suzuki, J. Phys. Soc.
Jpn. 74, 1674 (2005).
\bibitem{satom} M. Sato, Y. Takahashi and S. Fujimoto, Phys. Rev.
B 82, 134521 (2010).
\bibitem{goldman} N. Goldman, I. Satija, P. Nikolic, A. Bermudez,
M. A. Martin-Delgado, M. Lewenstein and I. B. Spielman, Phys. Rev.
Lett. 105, 255302 (2010).
\bibitem{lewenstein} A. Kubasiak, P. Massignan and M. Lewenstein,
Europhys. Lett. 92, 46004 (2010).



\end{thebibliography}
\end{document}